\documentclass[10pt,a4paper,twocolumn]{article}
\usepackage{lineno}
\usepackage{hyperref}
\modulolinenumbers[1]
\usepackage{authblk}
\usepackage{graphicx}
\usepackage[a4paper, total={6in, 9in}]{geometry}
\usepackage[toc,page, title, titletoc]{appendix}
\usepackage[symbol]{footmisc}

\begin{document}

\title{{Modeling of a MeV-scale Particle Detector Based on Organic Liquid Scintillator}}

\author[1,2]{Yu.M.~Malyshkin\footnote{*}}
%\ead{malyshkin@inr.ru, yu22mal@gmail.com}

\author[1,3]{A.M.~Gangapshev} % +BNO
\author[1]{V.N.~Gavrin}          % +BNO
\author[1]{T.V.~Ibragimova}      % BNO
\author[1]{M.M.~Kochkarov}       % BNO
\author[1]{V.V.~Kazalov}         % BNO
\author[1]{D.Yu.~Kudrin}
\author[1,3]{V.V.~Kuzminov}   % +BNO
\author[1]{B.K.~Lubsandorzhiev}
\author[1]{G.Ya.~Novikova}
\author[1,4]{V.B.~Petkov}       % BNO
\author[1]{A.Yu.~Sidorenkov}
\author[1]{N.A.~Ushakov}
\author[1]{E.P.~Veretenkin}
\author[1]{D.M.~Voronin}
\author[1]{E.A.~Yanovich}

\affil[1]{Institute for Nuclear Research RAS, Russia}
\affil[2]{National Institute of Nuclear Physics, Section of Roma Tre, Italy}
\affil[3]{Kabardino-Balkarian State University, Russia}
\affil[4]{Institute of Astronomy RAS, Russia}

\twocolumn[
  \begin{@twocolumnfalse}

  \maketitle

\begin{abstract}
The detectors based on the liquid scintillator (LS) monitored by an array of photo-multiplier
tubes (PMT) are often used in low energy experiments such as neutrino oscillation studies
and search for dark matter. 
Detectors of this kind operate in an energy range spanning from hundreds of keV to a few GeV 
providing a few percent resolution at energies above 1~MeV and allowing to observe fine 
spectral features. This article gives a brief overview of relevant physical processes 
and introduces a new universal simulation tool LSMC (Liquid Scintillator Monte Carlo) 
for simulation of LS-based detectors equipped with PMT arrays. 
This tool is based on the Geant4 framework and provides supplementing functionality 
for ease of configuration and comprehensive output. 
The usage of LSMC is illustrated by modeling and optimization of a compact detector 
prototype currently being built at Baksan Neutrino Observatory.
\end{abstract}
  \end{@twocolumnfalse}
  \vspace{1cm}
]

\tableofcontents

\footnotetext{* Corresponding author: malyshkin@inr.ru}

\section{Introdution}

Construction of new large neutrino detectors is necessary for solving a number of problems of 
modern neutrino physics, astrophysics and geophysics. In the past, large-volume 
unsegmented detectors have already demonstrated their capabilities, however, 
larger target masses, of kiloton scale, are needed for new generation experiments
in order to compensate the 
low strength of neutrino interactions and provide enough statistical significance
of results. 

A number of particle detection methods has been developed and different 
working substances were used as target. The organic liquid scintillators (LS), along with 
water, are the most suitable working substances to construct kiloton-scale neutrino 
detectors. In contrast to water Cherenkov detectors scintillation detectors are 
capable to reach energies down to hundreds keV, and thereby allow to detect, for example, 
geo-neutrinos and solar neutrinos from CNO cycle of sub-MeV energies.

The scintillation technique is common for an entire class of detectors which are 
now working in the field of neutrino physics, astrophysics and geophysics 
(KamLAND~\cite{KamLAND_2003}, Daya Bay~\cite{DYB_2018}, Double Chooz~\cite{DoubleChooz_2012}, 
RENO~\cite{RENO_2018}, Borexino~\cite{Borexino_2015} and others).
Double Chooz, Daya Bay and RENO have discovered the non-zero value of the $\theta_{13}$ 
oscillation parameter and Daya Bay later provided its most accurate estimate.
KamLAND and Borexino have found the first evidence for the weak geo-neutrino signal 
originating from radioactive elements embedded in the Earth's crust. 

The techniques developed for these experiments keep improving and will be used for 
a number of new LS-based detectors being actively planned 
(LENA~\cite{LENA_2012}, Theia~\cite{THEIA_2018}) and already being constructed 
(JUNO~\cite{JUNO}, Jinping~\cite{Jinping_2017}) along with existing 
facilities being upgraded (SNO+~\cite{SNO+_2016}, RENO-50~\cite{RENO-50_2015}). 
A new large-volume scintillation detector 
at the Baksan Neutrino Observatory for purposes of neutrino geo- and astrophysics has been
discussed for a long time. Recently resumed R\&D activities are aimed at the creation of
such a detector \cite{Barabanov2017}. A small scale prototype is 
already under construction.

Computer modeling of LS-based detectors plays an important role in their 
designing, understanding of their operation and event reconstruction. One can use an
empirical model giving a prediction of the observed energy based on initial energy, 
energy deposition in the scintillating volume and position of scintillation~\cite{Lebanowski_2018}. 
This approach requires a number of assumptions such as Guassian distribution of detector 
response and small contribution of Cherenkov light. However, the assumptions made 
in~\cite{Lebanowski_2018} often hold in practice. We should also note that the 
determination of deposited energy sometimes requires a dedicated simulation. 

In practice Monte Carlo (MC) simulations are usually used for this task giving more 
flexibility at a price of heavier computations. Simple approximated approaches are used along with more 
comprehensive computer programs often based on existing frameworks. In any case developing of such a tool takes 
significant effort and time. Future experiments may benefit from using a universal MC tool having 
enough flexibility and functionality.

Such a tool called LSMC (Liquid Scintillator Monte Carlo) has been created with the aim to 
help quickly start simulations. It is developed on top of the Geant4 
toolkit~\cite{Geant4_2003, Geant4_2006} inheriting its physical models, functionality and philosophy.
It is designed to be configurable for a wide range of detector geometries, 
materials and PMT models with no need of any coding. Instead it is fully controlled 
via interactive commands or macro files. Another goal of this project 
is to improve existing models based on the experience gained in real applications, 
e.g.\ with the upcoming data from the detector prototype in Baksan.

An overview of the relevant physical processes and detection techniques is 
given in Section~\ref{sec:models}.
Section~\ref{sec:lsmc} describes the LSMC tool and its validation. 
Section~\ref{sec:prototype} gives an example of LSMC application for detector design.
Finally, in Section~\ref{sec:conclusions}, we summarize our results and present our conclusions.

\section{Light production and detection in liquid scintillators}
\label{sec:models}

\subsection{Scintillators}

Organic liquid scintillators consist of a solvent medium and a small amount 
of fluor, often with an addition of a tiny amount of wavelength shifter. The 
kinetic energy of charged particles crossing the medium is mainly deposited in 
the solvent and excites its molecules. The excitation energy is rapidly 
transferred to the fluor, usually by a non-radiative dipole-dipole interaction. 
If the emission spectrum of a fluor has a significant overlap with its own absorption 
spectrum or the one of the solvent, a wavelength shifter is needed for re-emission 
to a higher wavelength region.

The choice of a specific LS mixture depends on the physics goals and the
technical design of the detector. The scintillator performance is usually 
optimized by maximizing its light yield, absorption and scattering lengths, 
and minimizing its scintillation decay time. Besides that, the emission spectrum 
should be in the region of the maximal sensitivity of the photo-multiplier tubes (PMTs).
It makes a direct impact on the detector performance, i.e.\ the energy and time 
resolution, and the low energy threshold. 

Often the neutrino signal can not dominate the background even if the target mass
is extremely large. When this is the case, the low-background conditions become crucial and ``active'' background suppression is to be used, e.g.\ event selection 
based on fast coincidences, rejection of events from the outer part of detector (fiducialization),
time veto after energetic muon passing the detector, discrimination of particle type by 
different emission time profile (pulse shape discrimination) and others. The final 
effectiveness of background suppression depends on a number of LS features such as 
the light yield, the fluorescence time profile, the transparency and the 
radiopurity.
These quantities are mainly defined by the purity of the original materials, 
as well as by temperature and aging effects.

The light yield of LS depends greatly on the concentration of fluor(s) added 
to the solvent, increasing at low concentrations and then saturating at a certain level. 
So, the optimization of 
the amount of collected light with regard to the self-absorption effects is 
needed for each specific LS mixture. 

In the last decades large progress has been made in the study of the properties 
of different LS mixtures. At present linear alkylbenzene (LAB) is one of the 
most popular solvents used to produce LS for large-scale detectors. Development 
of a large scintillation detector requires to consider also its cost and safety. 
In this respect, a scintillator on the basis of LAB has indisputable advantages, 
such as hypotoxicity, low volatility, and high flash point. Besides, LAB is a 
low-cost compound produced by the industry in large amounts.

\subsection{Photosensors}

Photomultiplier tubes (PMTs) are the most commonly used photon detectors in 
nuclear and particle physics experiments, in particular in astroparticle experiments.
For large-scale LS detectors the use of large area PMTs is a must in order to 
collect as much light as possible, while smaller detectors can be equipped 
with silicon photo-multipliers (SiPM). Here only the former ones are considered. 

In PMTs an individual photon can be registered via an electron produced by the 
photo-electric effect taking place in its front surface (photo-cathode) and then 
captured by electromagnetic field and amplified by a cascade of dynodes to a microscopic 
pulse. PMT performance affects the main parameters of the whole detector. For instance, the PMT 
sensitivity to scintillation light largely defines the sensitivity of LS detector as 
a whole. The other parameters of PMTs such as single photo-electron response, timing, 
dark current as well as the stability of these parameters are of crucial importance too.

The sensitivity of PMTs is characterized by photon detection efficiency (PDE). 
This parameter is a product of quantum efficiency of PMT's photo-cathode (QE), 
collection efficiency of photo-electrons (CE) and 
probability to detect photo-electron by PMT's dynode system (PD):
\begin{equation}
  {\rm PDE} = {\rm QE}(\lambda) \times {\rm CE}(\theta) \times {\rm PD}(V),
\end{equation}
where QE depends on photon wave-length $\lambda$, CE depends on incident azimuthal position $\theta$ 
and PD depends on photo-cathode to first dynode voltage $V$.

Modern large area PMTs have high quantum efficiency, e.g.\ the so-called super-bialkali 
photo-cathodes allow to reach PDE of more than 35\% at wavelengths of 360--400 nm. 
On top of that they have good single photo-electron response: the typical resolution of the single 
photo-electron peak in their photo-electron ADC charge distribution is 30-40\% at gain 
of $10^6$--$10^7$. This peak is clearly separated from the pedestal at lower ADC channels 
with peak-to-value ratio of more than 2.5--3. The better single photo-electron 
response, the higher value of PD. For modern large area PMTs the value of PD is close to 100\%.

PMTs register noise pulses, known as dark current, present even in complete darkness. It is 
described by the dark current counting rate above certain threshold 
which is usually set to a fraction of single photo-electron, e.g.\ a quarter of mean value of 
photo-electron charge (0.25 p.e.). Typical values for dark current counting rate of large area 
PMTs are of the order of $10$ kHz at room temperature. 

PMT timing is characterized by single photo-electrons transit time spread (TTS or jitter). 
Typical values of TTS of large PMTs are in the range of 3--5~ns (FWHM). One should take 
into account other effects coming from the coupled electronics and distorting PMT pulse shape.

\section{LSMC}
\label{sec:lsmc}

\subsection{LSMC introduction}

A dedicated software called Liquid Scintillator Monte Carlo (LSMC) has been 
developed for modeling of detectors based on LS and equipped with PMT arrays. 
LSMC is based on the Geant4 toolkit~\cite{Geant4_2003, Geant4_2006} utilizing its physical models, 
user interface and visualization functionality. 

LSMC is designed to provide a user an easy way to configure own detector parameters 
for simulation via interactive commands or by usage of macro files, and allows to 
configure a variety of setups with no need to alter the code. A typical detector 
like the ones considered in Sections~\ref{ssec:dyb} and~\ref{ssec:pr_geom} is defined by $\sim100$~commands.
In contrast, building an application using pure Geant4 would require an implementation
of many objects using C++ classes provided in the package. 

For the event generator LSMC utilizes Geant4 General Particle Source~\cite{Geant4_GPS}, which 
provides a rich functionality for setting primary particle source: 
particle type, energy, position, angular distributions etc. 

All the standard Geant4 materials~\cite{Geant4_MDB} are available in LSMC. Some other 
materials often used in construction of LS-based detectors, such as different scintillators, 
acrylic and mineral oil, are additionally implemented in LSMC.

Light emission is modeled with the Geant4 optical physics models (see 
``ElectromagneticInteractions/Optical Photons'' in~\cite{Geant4_PRM}). LS material emits 
photons according to a pre-defined or a user-defined scintillation yield and spectrum. 
The simulation includes quenching described by the Birks' law~\cite{Birks} and production 
of Cherenkov light~\cite{Cherenkov}. Although only a tiny fraction of photons in the 
region of PMT sensitivity is produced by Cherenkov effect, it makes an impact on the 
detector response non-linearity and is enabled in LSMC by default. 

Photons reaching PMT photo-cathode are counted with the probability defined by the QE 
curve. Different simple PMT shapes are available in LSMC: cylinder, half-sphere and 
``bulb'' (cylinder + sphere). The sizes can be set by user. Besides that LSMC provides 
three realistic PMT models with their QE curves: ETL-9351, Hamamatsu R5912 and 
Hamamatsu R7081-100. Optionally, a user can define own QE curves. The list of available 
PMT models is to be extended in future. 

Dark noise, gain fluctuations and other PMT processes are not modeled by LSMC and 
may be implemented in the future updates. At the moment they can be applied manually on 
top of the LSMC output. 

LSMC is to be publicly released in near future. However, 
it can be shared privately upon request. Interested groups are invited to contact the 
corresponding author.

\subsection{Input}

LSMC follows the concept of hierarchical organization of commands utilized in Geant4. 
The commands can be typed in interactive mode or written sequentially in macro files. 
The top level branches are responsible for configuration of event generator, 
physical processes, geometry, material, level of details for the output, 
visualization, as well as for run control. Most of them are the standard Geant4 commands.
An integrated help system provides a description for each command, which is a part of 
the Geant4 package.

A special top level branch '{\texttt /lsmc}' is added to group the LSMC-specific commands. 
The LSMC-specific commands include configuration of sizes and materials of the 
outer tank, the vessel for scintillator and auxiliary components. 
For the scintillator one can set the light output and the value of the Birks' constant. 
Special commands are available for choosing PMT type, optionally provide a specific QE
curve, and define PMT arrangement.

An example of an input macro file is provided in~\ref{ap:input}.

\subsection{Output}

The output of simulation is saved to a ROOT file \cite{ROOT_1997}.
The output data includes true event generator information (initial energy, 
position and momentum), energy deposition in different detector components, 
total number of photons, number of registered photo-electrons, number of fired PMTs, 
number of absorbed photons in detector components, and other information.
One can also enable the extended output mode, which triggers 
storage of information about each registered photo-electron, such as PMT ID, 
generation and absorption channel and photo-cathode hit time.

See~\ref{ap:output} for the full output description.

\subsection{Benchmarking with Daya Bay detector}
\label{ssec:dyb}

For verification of the correctness of the simulations performed by LSMC a 
benchmarked simulation has been carried out. Daya Bay detector has been 
chosen due to availability of published characteristics. The Daya Bay
detector system~\cite{DYB_2016_detector} consists of eight identical detectors 
immersed into three water pools and covered by muon detectors. 
The eight detectors are referred to as anti-neutrino detectors (AD). Each one 
consists of the inner vessel with Gd-doped LAB-based LS nested in another 
vessel filled with undoped LS. The outer vessel and an array of 192 8-inch PMTs 
(Hamamatsu R5912) are placed in a tank with mineral oil. The inner and outer 
vessels with LS are made of acrylic. One can refer to~\cite{DYB_2016_detector}
for more details.

\begin{figure}[!htb]
  \centering
  \includegraphics[width=\columnwidth]{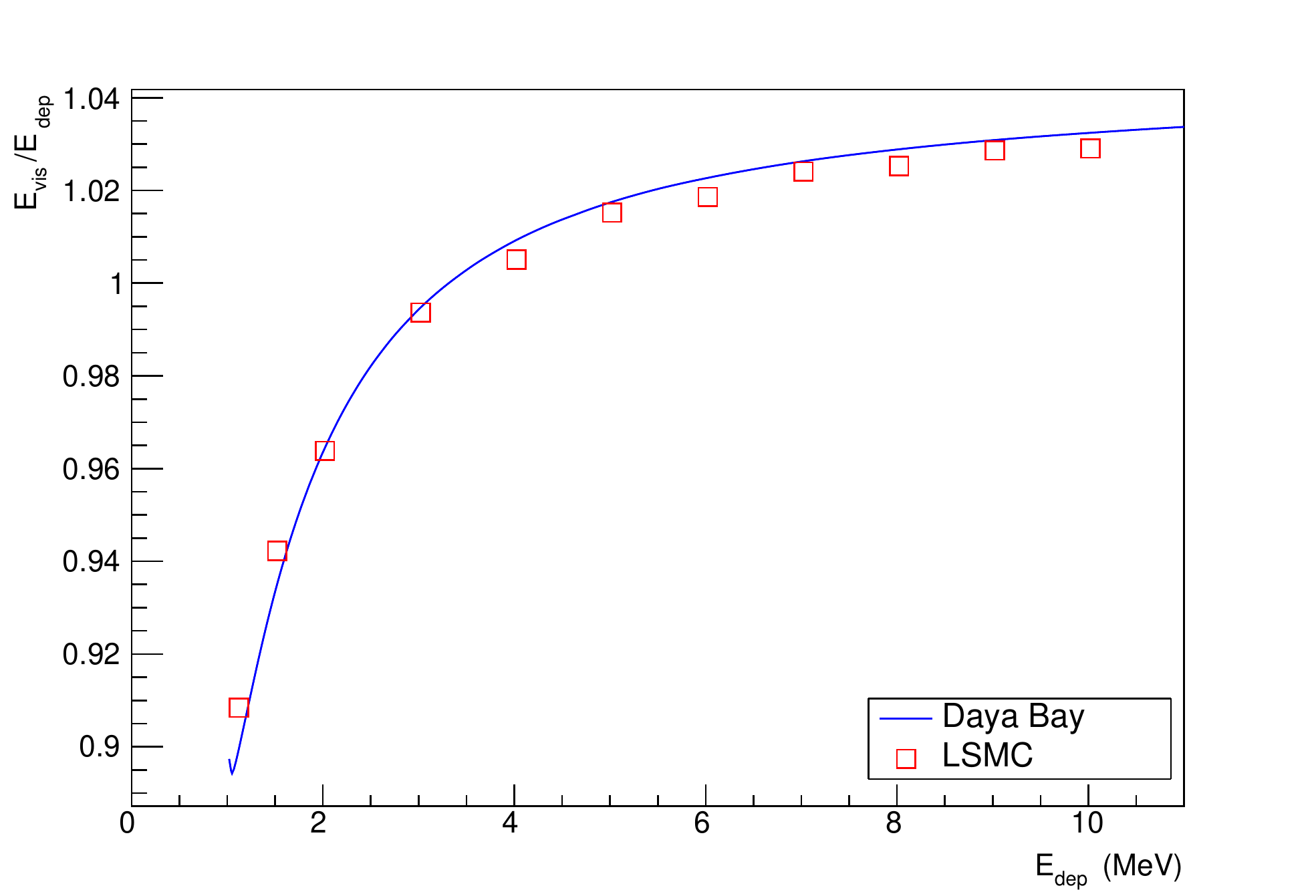}
  \caption{Daya Bay energy scale non-linearity $E_{\rm dep}/E_{\rm vis}$ for positrons 
  with the majority of electronics nonlinearity removed~\cite{DYB_2019_nl_data} (blue line), 
  and the result of a LSMC simulation with manually adjusted normalization (red squares).}
  \label{fig:dyb_nl}
\end{figure}

The scintillator light yield has been set to 9000 to match the normalization 
coefficient of $\sim170$ p.e./MeV
measured in Daya Bay. The Birks' coefficient has been set to $k_B=0.0158$~g/cm$^2$/MeV 
in accordance with~\cite{DYB_2019_nl}. The detector response non-linearity has been 
investigated for positrons in the energy range from 0 to 10 MeV, see Fig.~\ref{fig:dyb_nl}
showing dependence of $E_{\rm vis}/E_{\rm dep}$ on deposited energy $E_{\rm dep}$.
The visible energy $E_{\rm vis}$ corresponds to the detectable light and is proportional to 
the number of detected photo-electrons. For the case of positrons the deposited energy 
$E_{\rm dep}$ is almost always equal to the sum of kinetic energy and the energy of two 
annihilation gammas: $E_{\rm dep} = E_{\rm kin} + 1.022$ MeV. For the LSMC data the 
normalization coefficient of 168~p.e./MeV has been used. The results of LSMC simulation 
are compared with the corresponding non-linear energy model shown in Fig.~22 
of~\cite{DYB_2019_nl} (the majority of electronics non-linearities removed).

The agreement of the measurement-based non-linearity curve of the Daya Bay detectors 
with the LSMC prediction proves its reliability for simulation of similar detectors. 
It should be note, however, that one has carry about appropriate values for the 
scintillator light yield and Birks' constant. One should also remember about the 
contribution of Cherenkov light. To investigate it one can switch on and off the 
Cherenkov process in the simulation.

\section{Geo-neutrino detector prototype}
\label{sec:prototype}

Baksan Neutrino Observatory in Caucasus Mountains has facilities that are among the deepest 
in the world. The overburden of laboratories reaches 4700 m.w.e. 
One of them has been used for solar neutrino monitoring by the SAGE experiment \cite{SAGE_2009}.
The low reactor neutrino background makes this location very promising for 
geo-neutrino measurements. A kiloton-scale 
detector has been proposed in \cite{Domogatski_2004} for studies of Earth's crust. 
Such a detector may reinforce the future network of geo-neutrino monitors. 
In order to start-up this large project a small detector prototype has been 
designed and is currently under construction. 
LSMC was used at the designing stage. Some results are presented below 
to illustrate the functionality of LSMC.

\subsection{Baseline prototype geometry}
\label{ssec:pr_geom}

The prototype has about 420~kg of LAB-based scintillator with 
2,5-\hspace{0cm}Diphenyloxazole (PPO) and bisMSB admixtures.
The scintillator is contained within a sphere with the radius of 50~cm and varying 
thickness of 1-2~cm (in simulation fixed to 1~cm).
The sphere is made of acrylic, a transparent material with the refraction index similar 
to the one of LAB. It is put into a cylindrical tank filled with water
serving for protection from external radioactivity. The diameter and height 
of the tank are 240 and 280~cm respectively. Its dimensions are constrained 
by the requirement to fit the tunnel during the transportation underground.

An array of 20 10-inch Hamamatsu R7081-100 PMTs surrounds the scintillator 
contained in the acrylic sphere.
PMTs are placed at vertices of a regular dodecahedron at about 75~cm distance 
measured from the sphere center to PMT equators. The PMTs and the acrylic 
sphere are mounted on a stainless steel supporting structure. PMTs are 
also to be equipped with conical concentrators in order to increase the amount 
of gathered light.

\begin{figure*}[!htb]
  \centering
  \includegraphics[width=0.5\textwidth]{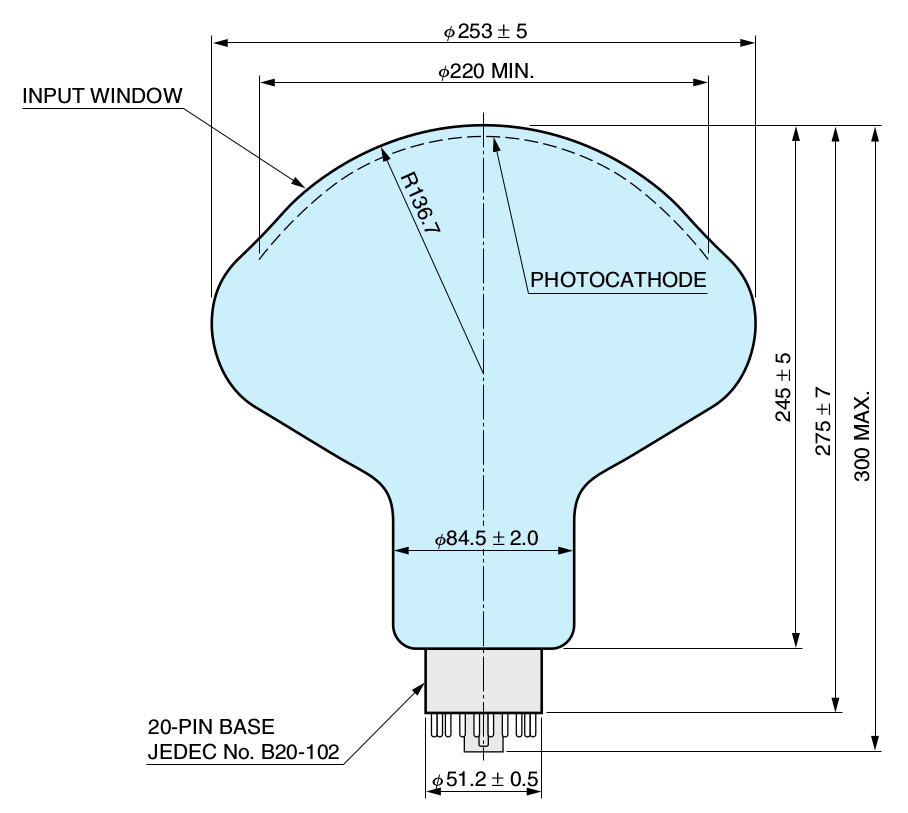}
  \includegraphics[width=0.46\textwidth]{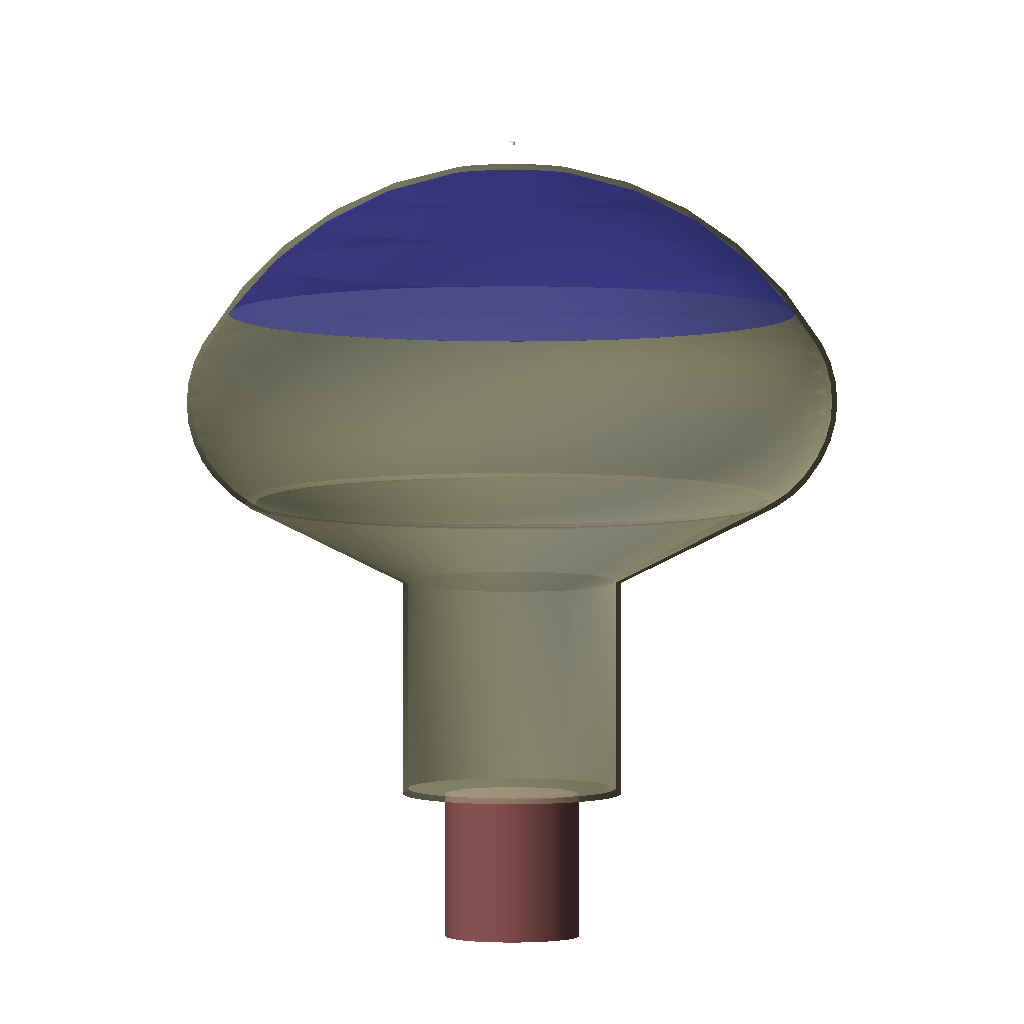}
  \caption{Hamamatsu R7081 PMT geometry: from Hamamatsu data sheets (left), LSMC model (right).}
  \label{fig:pmt_geo}
\end{figure*}

A realistic model of Hamamatsu R7081-100 tubes implemented in LSMC was used for 
this simulation. Its shape and dimensions taken from the manufacturer's data sheets 
are shown in Fig.~\ref{fig:pmt_geo}. 

\subsection{Modeling with LSMC}

The prototype has been modeled with the LSMC software in order to better 
understand the light absorption budget, impact of different geometry choices 
on light collection, and other aspects prior to the construction. In all 
simulations primary electrons with total energy of 1~MeV were generated 
uniformly over the entire liquid scintillator volume with isotropic 
directionality. Some smaller parts of the detector structures (like PMT holding structures, cables etc.) were omitted for
simplicity.

\subsection{LS and PMT configuration}

\begin{figure}[!htb]
  \centering
  \includegraphics[width=\columnwidth]{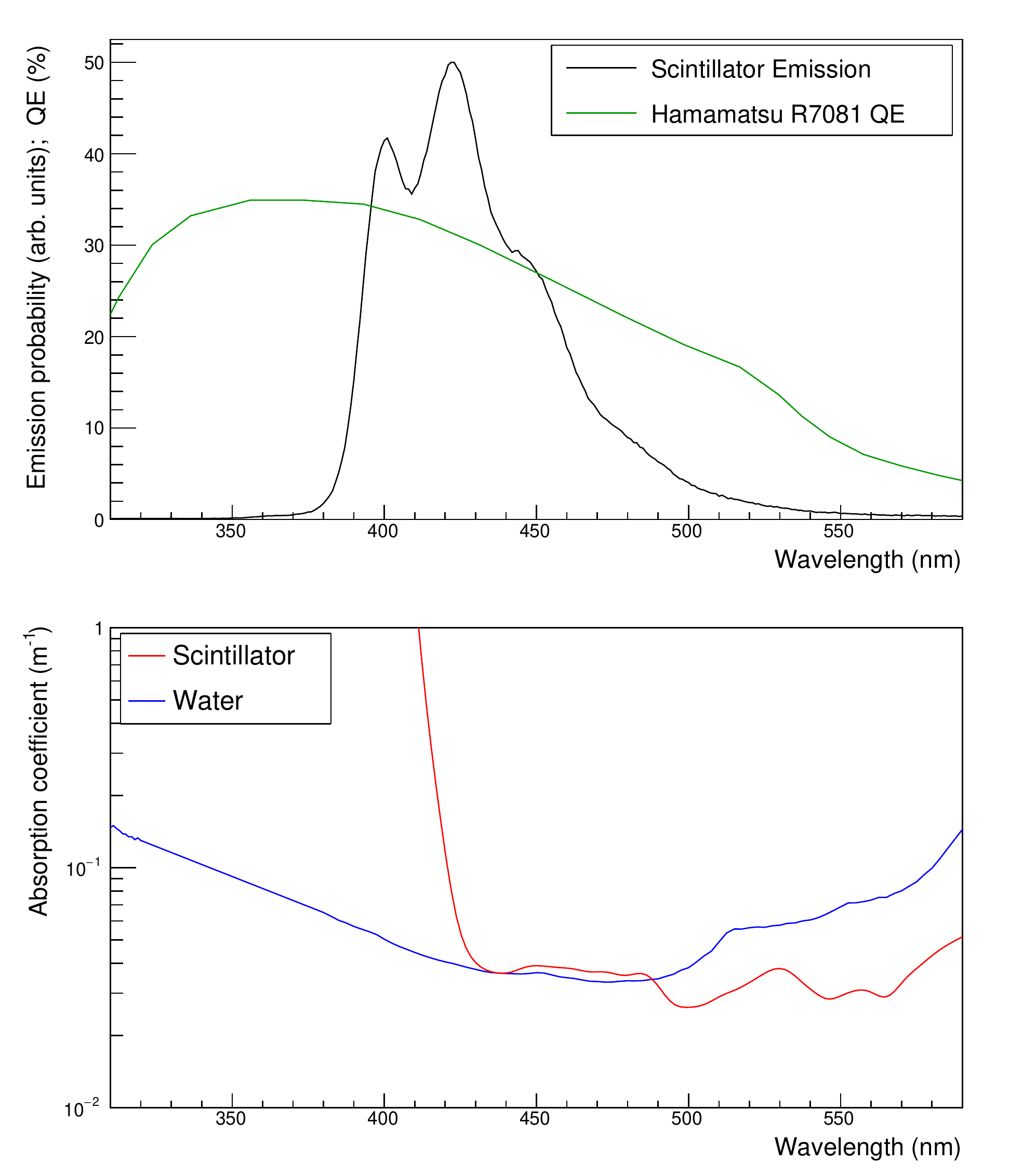}
  \caption{{\it Top panel:} Light emission curve (including re-emission) of LAB scintillation and PMT QE in LSMC.
           {\it Bottom panel:} light absorption spectra for LAB-based scintillator and water in LSMC. 
           The emission and absorbance data are taken from~\cite{Beriguete2014}; PMT QE curve is from Hamamatsu data sheets. }
  \label{fig:emi-abs_curves}
\end{figure}

The exact light yield of LAB and fluor concentrations are not determined yet, therefore an 
conservative estimate of 10000 photons/MeV (as in~\cite{Lebanowski_2018} and in~\cite{DYB_LAB_2010}) 
and an approximate emission spectrum are used for simulations. The emission spectrum 
is shown on the upper panel of Fig.~\ref{fig:emi-abs_curves} together with the QE 
curve of the photo-multiplier tubes to be used for the detector.
The bottom panel of Fig.~\ref{fig:emi-abs_curves} illustrates LAB and water absorption 
spectra used in simulation. More details on light emission and absorption in 
LAB-based scintillators can be found in~\cite{DYB_LAB_2010}. Here we used similar spectra.
The absorption length of water can vary in a wide range depending on the level of purification:
from several meters for ``tap'' water to a few hundreds of meters for ``ultra-pure'' 
water~\cite{SK_calib_2014, Sogandares1997, Pope1997}. Water transparency is not of crucial 
importance in our case due to the array size and we chose reasonable values 
from~\cite{Mobley1995, Querry1991} which could be reached easily for affordable price. 

\subsubsection{Geometrical configuration}

\begin{figure*}[!htb]
  \centering
  \includegraphics[width=0.31\textwidth, trim={3cm 0 3cm 0.5cm},clip]{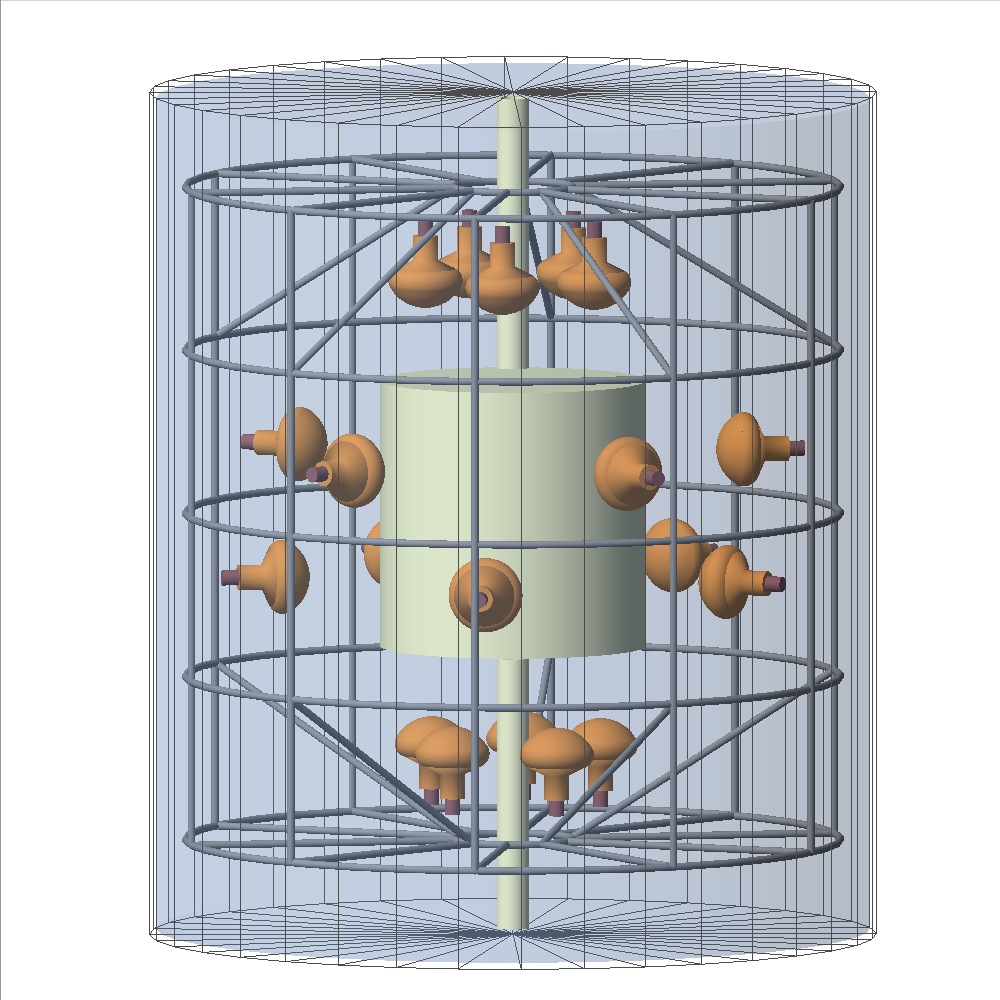}
  \includegraphics[width=0.31\textwidth, trim={3cm 0 3cm 0.5cm},clip]{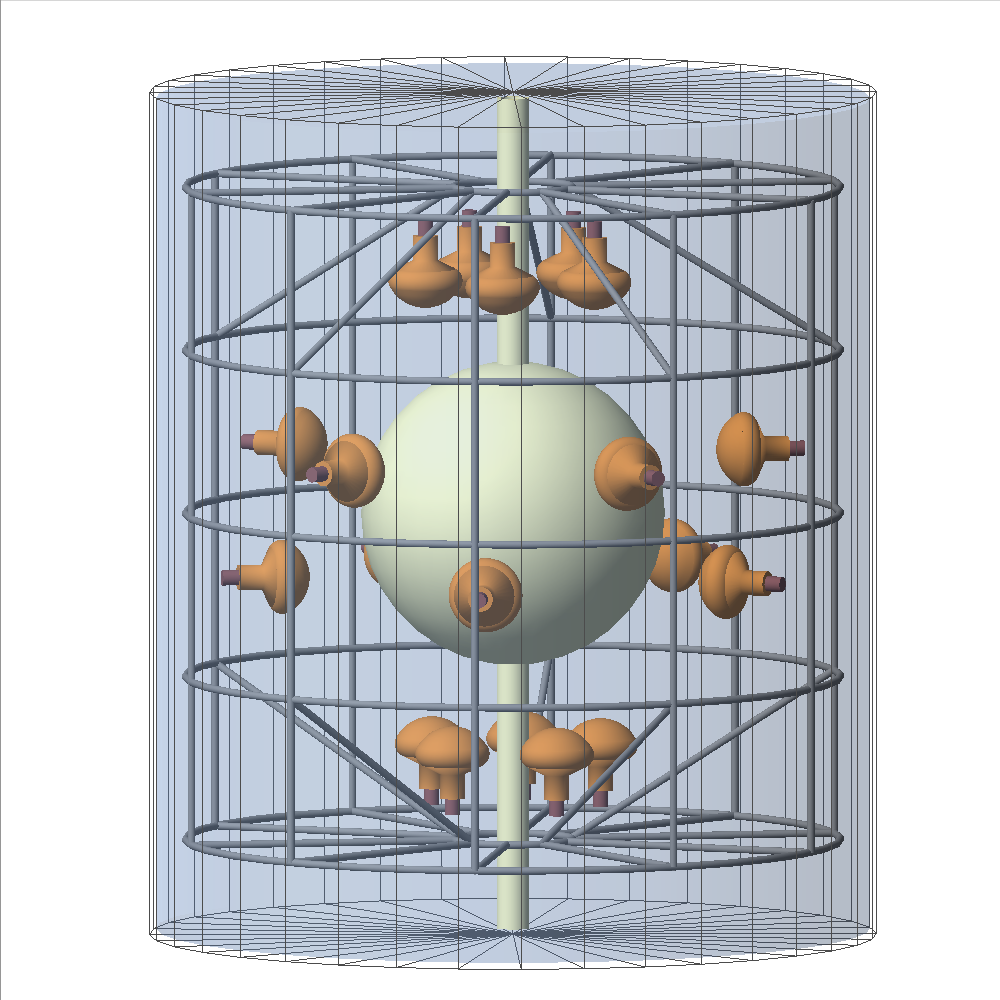}
  \includegraphics[width=0.31\textwidth, trim={3cm 0 3cm 0.5cm},clip]{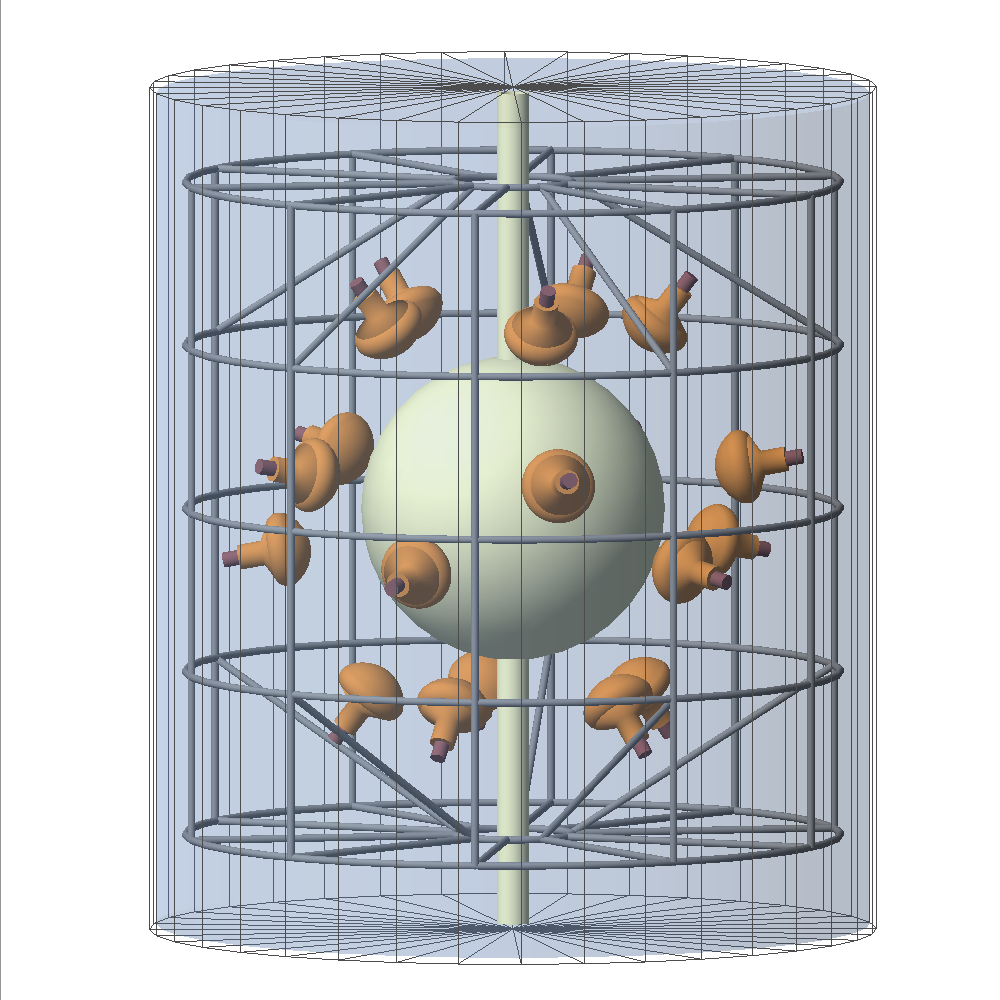}
  \caption{Different PMT arrangement options from left to right:
  1) cylindrical LS volume, cylindrical PMT arrangement (``cyl.-cyl.''),
  2) spherical LS volume, cylindrical PMT arrangement (``sph.-cyl.''),
  3) spherical LS volume, spherical PMT arrangement (``sph.-sph.'').}
  \label{fig:opt_geo}
\end{figure*}

The two most natural and commonly used geometry choices for LS vessel are 
cylinder and sphere. 
The first one is easier to fabricate, the other one provides better symmetry,
which in turn simplifies energy reconstruction. 
Here we compare the two options assuming the same LS mass of about 420~kg. 
The sphere has the inner radius of 49~cm, and the cylinder has the inner radius 
of 42.8~cm and the height of 85.6~cm.
These two options are considered below in combination with two PMT arrangement 
configurations: cylindrical and spherical, see Fig.~\ref{fig:opt_geo} for 
illustration.

\begin{figure}[!htb]
  \centering
  \includegraphics[width=\columnwidth]{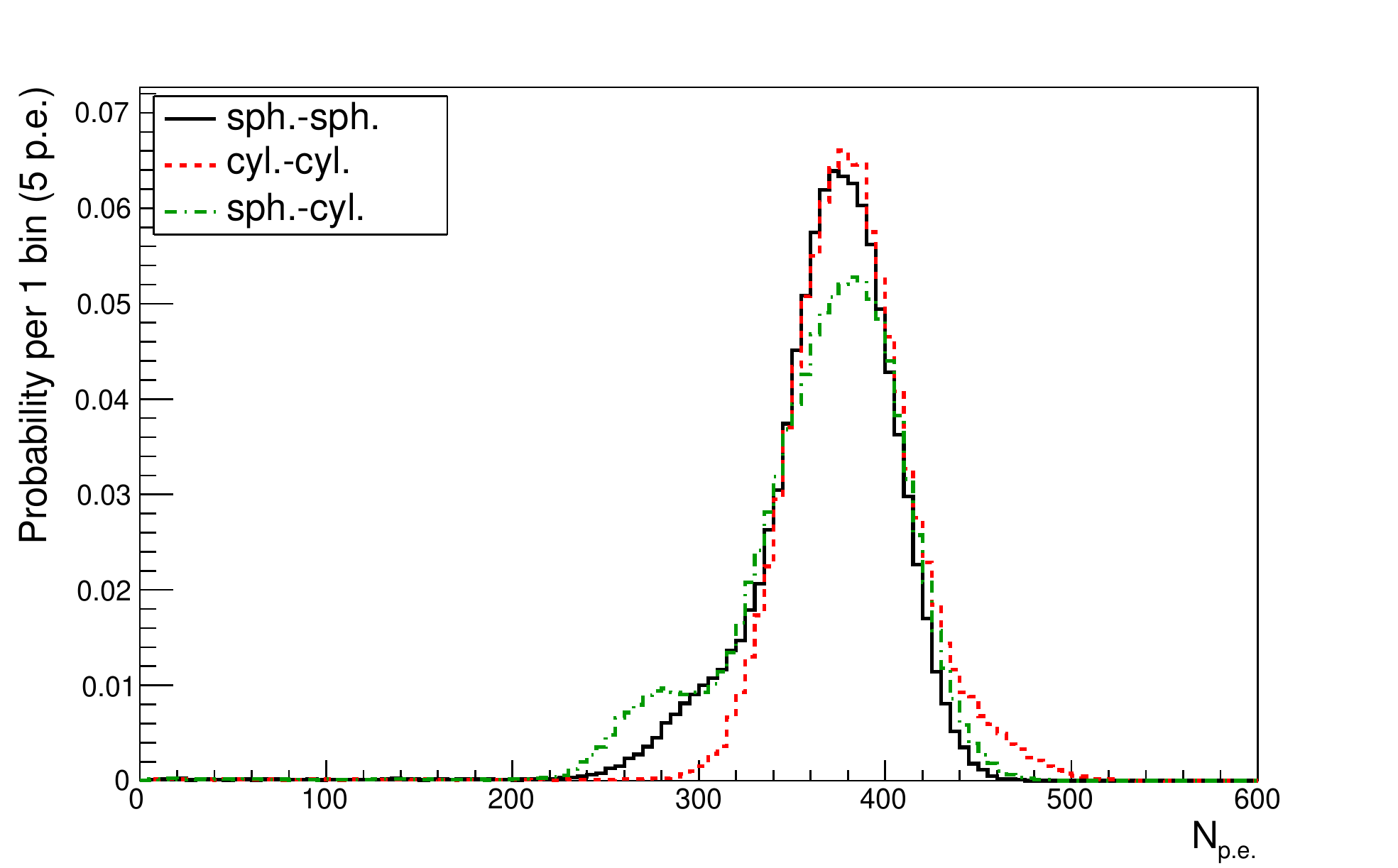}
  \caption{Detector response to 1~MeV electrons uniformly distributed over the LS volume
  for three arrangement options.}
  \label{fig:opt_resp}
\end{figure}

Fig.~\ref{fig:opt_resp} shows detector response in terms of number of 
registered photo-electrons for the three introduced combinations. 
In case of the ``shp.-sph.'' configuration (with spherical LS volume and 
spherical PMT arrangement) it consists of a peak with a shoulder. 
However, as it will 
be discussed later, the events contributing to the shoulder can be 
easily separated from the ones from the main peak. Thus, after event reconstruction, this feature will 
not spoil the final energy resolution.

The ``cyl.-cyl.'' configuration 
provides slightly better light collection than the other options,
but the peak smearing is a bit larger compared to the main peak 
of the ``sph.-sph.'' option. 
The mixed configuration with spherical LS volume and cylindrical PMT
arrangement (``sph.-cyl'') inherits the drawbacks of both other and, therefore, 
is the least favorable. 
The ``sph.-sph.'' option has been selected for the discussed here 
detector prototype and is considered below as the default. 

\subsubsection{Detector response structure}

\begin{figure}[!htb]
  \centering
  \includegraphics[width=\columnwidth]{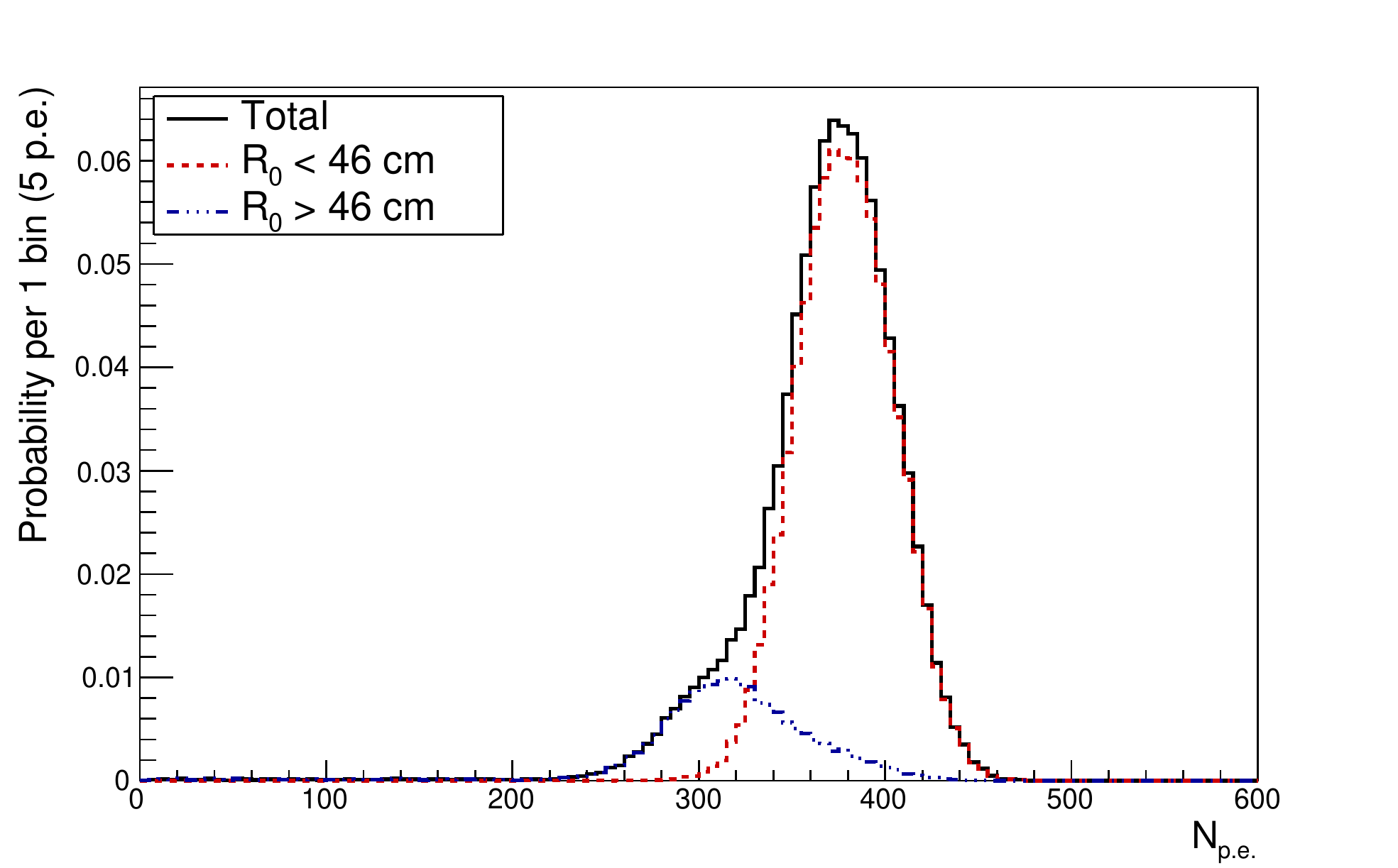}
  \caption{Detector response to 1~MeV electrons uniformly distributed over the LS volume.
  Contributions of the events originating from $R_0<46$~cm and $R_0>46$~cm are shown
  by red and blue lines respectively. }
  \label{fig:npe_bl}
\end{figure}

The photo-electron spectrum for the baseline configuration is shown in 
Fig.~\ref{fig:npe_bl}. The main peak corresponds to the events originating from 
the inner part of the scintillator sphere, while the smaller one is formed 
by the events from the outer zone. The red and blue lines show the corresponding
contributions. Photons emitted in vicinity of the acrylic sphere ($R>46$~cm in 
our case) may have the angle of incidence larger than the critical one 
when they reach the acrylic-water boundary. These photons are reflected back to 
the LS volume. Within a spherical volume they have the same angle of incidence 
when they reach acrylic boundary again (neglecting scattering), and can never 
penetrate to water. This portion of photons is not detected and such events 
with partially lost luminosity form the smaller peak of the response spectrum.

\subsubsection{Absorption channels}

Table~\ref{tab:abs_channels} shows the photon absorption budget for the 
baseline configuration and two modified detector configurations: 1) with 
black inner surface of the outer tank and 2) with light concentrators mounted on PMTs. 
In particular one can 
see that the most of the light is absorbed by LS, so even for 
such a small detector as considered here the LS transparency 
plays an important role. 

\begin{table*}[!htb]
  \centering
  \caption{Share of absorption channels given in percents for the baseline setup, and for two modified configurations.}
  \label{tab:abs_channels}
  \begin{tabular}{l|ccc}
    \hline\hline
    Absorbing                & Baseline       & Baseline with  & Baseline with \\
    material                 & configuration  & black tank     & concentrators  \\
    \hline
    Scintillator             & 35             & 33               & 36     \\
    Acrylic sphere           & 4              & 3                & 5      \\
    Water                    & 8              & 3                & 7      \\
    Tank                     & 18             & 49               & 12     \\
    PMT glass and coating    & 0.1            & $<$0.1           & 0.1    \\
    Photo-cathode            & 12             & 7                & 20     \\
    Other components         & 23             & 6                & 20     \\
    \hline\hline
  \end{tabular}
\end{table*}

\begin{figure}[!htb]
  \centering
  \includegraphics[width=\columnwidth]{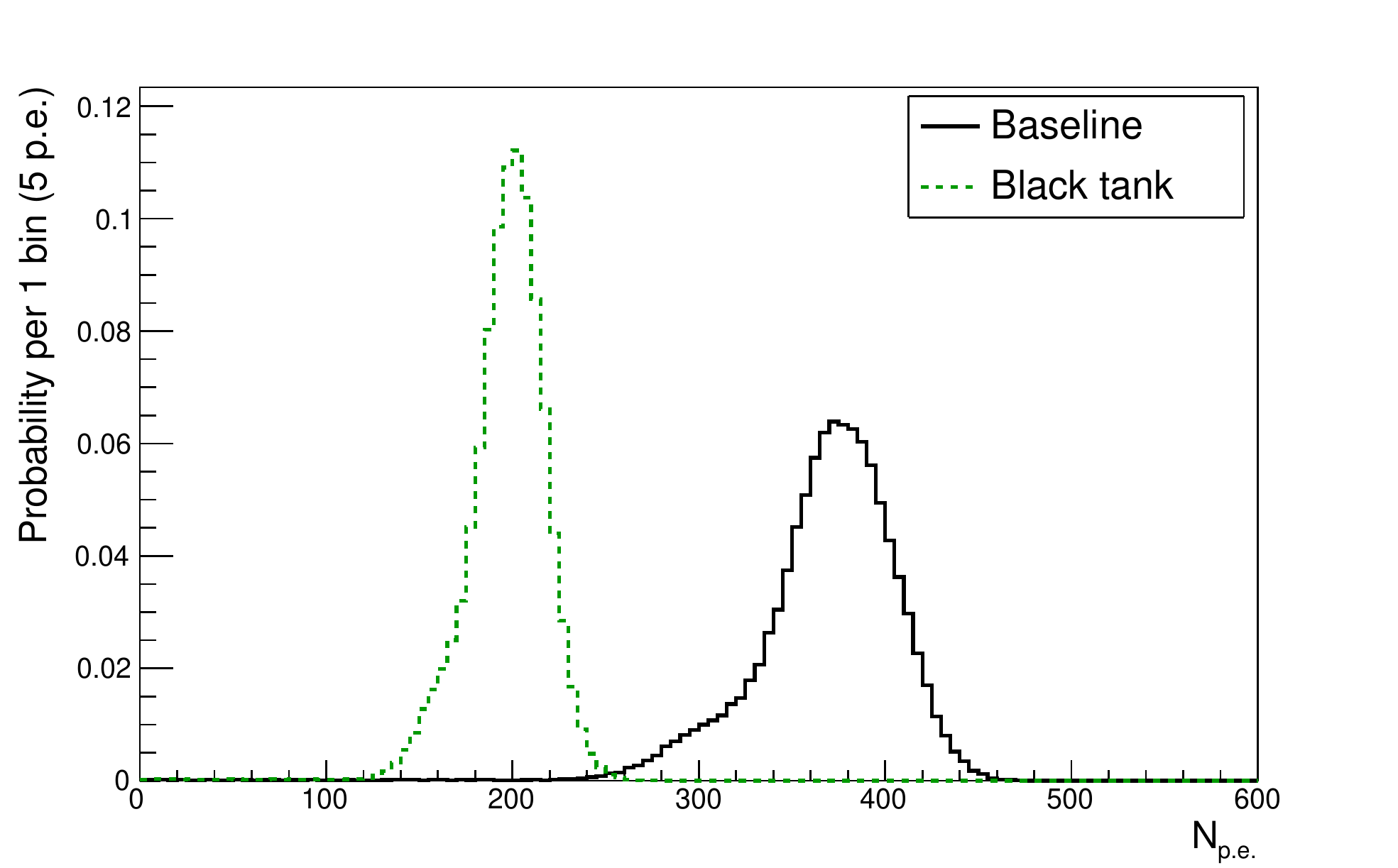}
  \caption{Detector response to 1~MeV electrons uniformly distributed over the LS volume 
  for 85\% (baseline) and 0\% (black tank) reflectivity of the inner surface of the water tank.}
  \label{fig:black_tank}
\end{figure}

The default reflectivity of the water tank inner surface is 85\%, and a non-negligible 
fraction of light can be reflected, and then reach a PMT photo-cathode. From one side 
the reflected photons increase the overall light collection. From the other side they
have longer paths, which leads to additional stochastic effects and spoils the energy 
resolution. One can consider the inner surface of the water tank having zero 
reflectivity. In this case (see 3rd column of Table~\ref{tab:abs_channels}) the amount of light absorbed 
on its inner surface increases by factor of about 2.7 with respect to the default 
configuration. As the result the total light collection (absorption on photo-cathode) decreases 
by a factor of about 1.7. Fig.~\ref{fig:black_tank} shows a comparison of photo-electron 
spectra with 85\% (default) and 0\% reflectivity of the tank inner surface. 

The other components, mainly supporting structures, may also play an important 
role in the photon collection budget. They are considered here totally 
non-reflecting and absorb from 6\% to 23\% of photons in the considered 
configurations.

\subsubsection{Concentrators}
\begin{figure*}[!htb]
  \centering
  \includegraphics[width=0.65\textwidth]{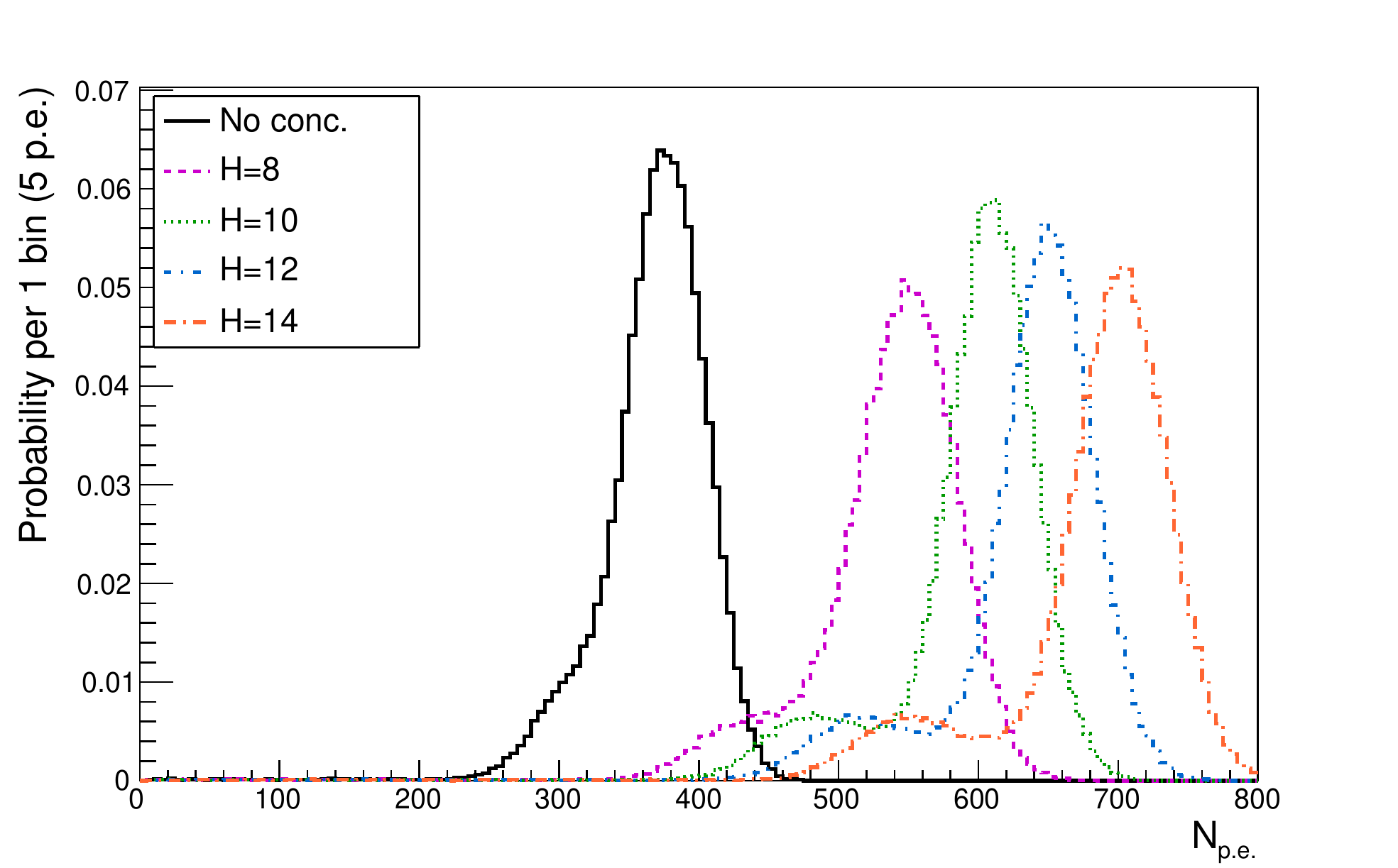}
  \includegraphics[width=0.33\textwidth, trim={3cm 0.5cm 3cm 0.5cm},clip]{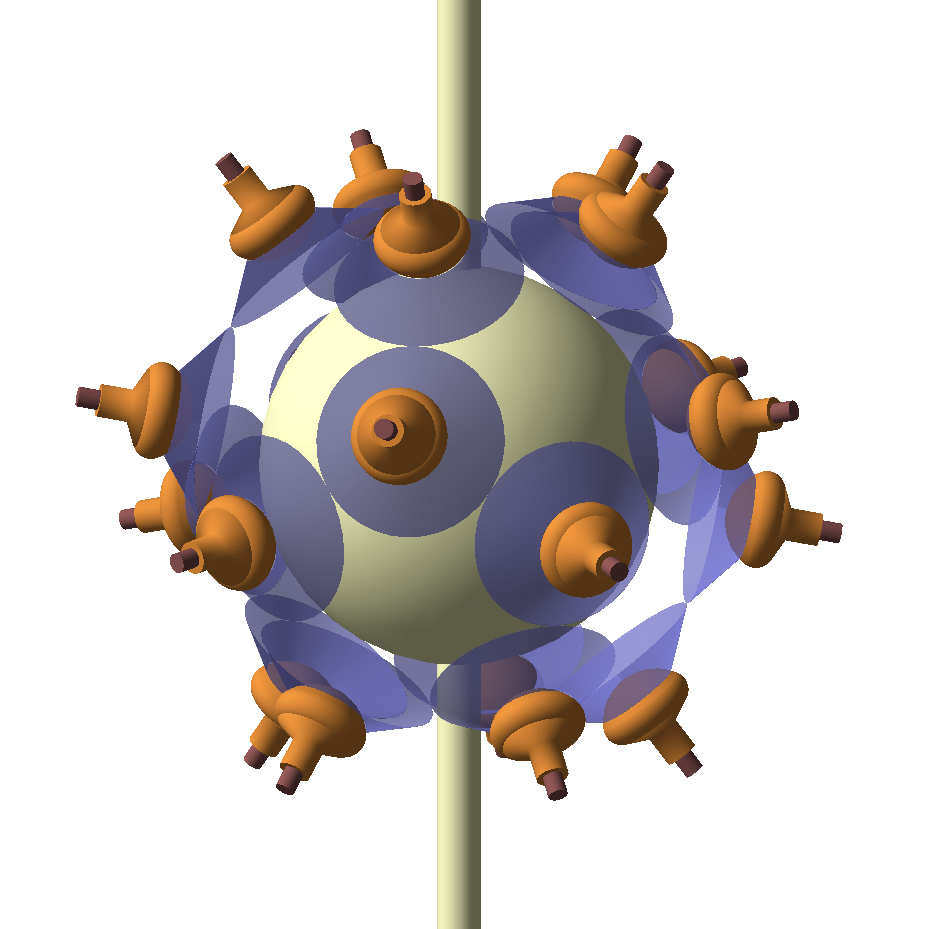}
  \caption{Light concentrators of 11~cm height mounted on PMTs (right side) and 
  response functions for concentrators of 8, 10, 12 and 14~cm height shown by color lines. 
  Detector response with no concentrator is shown by black line for comparison. }
  \label{fig:conc}
\end{figure*}

The light collection efficiency can be increased by equipping PMTs with concentrators. 
This idea has been already used in other neutrino detectors, e.g.\ in 
Borexino~\cite{Borexino_conc}. For the best performance the shape of the concentrators 
must be properly calculated. Here for simplicity we consider truncated cone shape, 
see the right side of Fig.~\ref{fig:conc} for illustration.  
We optimized the cone dimensions using LSMC simulation, however understanding its limitation. 
Assuming that the larger radius must be as large as possible to capture more light, 
and the smaller one must coincide with the photo-cathode edge, the only parameter 
for optimization is the height of the truncated cone.

A set of calculations with different concentrator heights has been performed in order 
to find the optimal one. The corresponding response functions for some of them
are shown in Fig.~\ref{fig:conc}. The one with the height of 10~cm 
provides the best performance in terms of resolution. Therefore, we conclude 
that the concentrators used for the prototype should have a height of about 9--11~cm.
The corresponding photon absorption budget (for the concentrator height of 10~cm) is 
presented in the 4th column of Table~\ref{tab:abs_channels}.

\section{Conclusions}
\label{sec:conclusions}

Modern simulation techniques allow us to construct realistic models and perform
fine tuning for detectors at the early designing stage. One can learn a lot about the 
impact of different geometry and material choices from such simulations. Existing
general purpose software and libraries are available but significant effort is
still needed to develop a ready to use computer program for a particular use case. 
Specialized tools for specific detector classes may help to start simulations quicker.

Introduced here LSMC software allows to model LS-based detectors with no need 
of coding, while offering a rich functionality. It includes geometry and material 
configuration, type of PMTs and different ways of their arrangement. 
LSMC provides a large set of output information in the widely used ROOT-format.

LSMC uses the state-of-the-art libraries for physical processes from the Geant4 
package. Thus, any updates in new versions of Geant4 automatically propagate to 
LSMC.
From the other side real data measured and benchmarked with LSMC (e.g.\ from the new
detector prototype at Baksan) may 
help to spot any problems, especially for the optical processes, and provide a
useful input for the Geant4 developers and users. 

\section{Acknowledgements}
The simulated detector prototype configuration
is based on the scientific equipment of UNU GGNT BNO INR RAS partially
acquired with financial support of the Ministry of Science and Higher
Education of the Russian Federation: agreement N 14.619.21.0009, unique
identifier of the project RFMEFI61917X0009.

The work was supported by the Russian Science Foundation, project no.
17-12-01331.

\onecolumn
\bibliography{nuphys}{}

\clearpage

\onecolumn
\begin{appendices}

\section{Example of input macro file for LSMC}
\label{ap:input}

An example of simulation configuration (for Baksan prototype):

\small
\begin{verbatim}
 ----------------------------------------------------------------------------------------- 
# PMT configuration
/lsmc/pmt/type R7081-100
/lsmc/pmt/shape R7081
#
# PMT arrangement
#                       R   theta   N phi0
/lsmc/pmt/addRingRTheta 79 37.3758  5 18
/lsmc/pmt/addRingRTheta 79 79.1874  5 18
/lsmc/pmt/addRingRTheta 79 100.8126 5 54 
/lsmc/pmt/addRingRTheta 79 142.6242 5 54
#
# Concentrators
/lsmc/enableConc 24.4 11.0 # R H (in cm)>  
#
# Scintillator
/lsmc/scint/material LAB
/lsmc/scint/shape sphere     
/lsmc/scint/radius 49. cm
/lsmc/scint/holderThickness 2. cm
/lsmc/scint/holderMaterial Acrylic
/lsmc/scint/yield 10000
#
# Inner vessel configuration
/lsmc/tank/radius 120 cm
/lsmc/tank/height 280 cm
/lsmc/tank/thickness 2.0 cm
/lsmc/tank/reflectivity 0.85
/lsmc/tank/fillingMaterial Water
#
# Additional structures
# LS tubes
/lsmc/misc/addTube ls_tube 4.5 5.5 Acrylic 0 0 49   0 0 140
/lsmc/misc/addTube ls_tube 4.5 5.5 Acrylic 0 0 -49  0 0 -140
# vertical tubes                                 x1   y1   z1     x2   y2   z2
/lsmc/misc/addTube truss 0 1.25 StainlessSteel  110    0  110    110    0 -110
...
# horizontal tubes, top                          x1   y1   z1     x2   y2   z2
/lsmc/misc/addTube truss 0 1.25 StainlessSteel  110    0  110     20    0  110
...
# horizontal tubes, bottom                       x1   y1   z1     x2   y2   z2
/lsmc/misc/addTube truss 0 1.25 StainlessSteel  110    0 -110     20    0 -110
...
# inclinated tubes, top                          x1   y1   z1     x2   y2   z2
/lsmc/misc/addTube truss 0 1.25 StainlessSteel   89   65   55     16   12  110
...
# inclinated tubes, bottom                       x1   y1   z1     x2   y2   z2
/lsmc/misc/addTube truss 0 1.25 StainlessSteel   89   65  -55     16   12 -110
...
# Ring-tubes, side
/lsmc/misc/addTorus truss 0 1.25 110 StainlessSteel 0 0  110
...
# Ring-tubes, top and bottom
/lsmc/misc/addTorus truss 0 1.25 20 StainlessSteel 0 0  110
/lsmc/misc/addTorus truss 0 1.25 20 StainlessSteel 0 0 -110
#
/run/initialize
#
# Primary particles
/gps/particle e-
/gps/energy 1 MeV
/gps/pos/type Volume
/gps/pos/shape Sphere
/gps/pos/radius 49 cm
/gps/ang/type iso
#
# Inactivate optical processes
# (Cerenkov, Scintillation, OpAbsorption, OpRayleigh, OpMieHG, OpBoundary, OpWLS)
#/process/inactivate Cerenkov
#/process/inactivate OpAbsorption
#/process/inactivate OpRayleigh
#
/run/printProgress 5000
/run/beamOn 100000 
 ----------------------------------------------------------------------------------------- 
\end{verbatim}
\normalsize

Symbol '\#' stands for comments, the lines starting with this character are ignored by LSMC.
To explore all the commands and descriptions one can run './LSMC' and type 'help', 
then walk through the menu to see all the options.

\section{Output structure of LSMC}
\label{ap:output}

The LSMC output is stored in ROOT format~\cite{ROOT_1997} and has two modes: normal and extended. 
The later one additionally includes photon-wise info and can be enables by command 
{\texttt /lsmc/recPhotonInfo True}. The data structure is the following (the data fields 
marked with ``{\texttt *}'' are only available in the extended output mode):

\small
\begin{verbatim}
 -----------------------------------------------------------------------------------------
  Data type   Field name              Description
 -----------------------------------------------------------------------------------------
  TTree       pmt_pos                 List of PMT positions in the detector. Index number
                                        in the list corresponds to its ID.
                pmtX_cm, pmtY_cm, pmtZ_cm
                                      x-, y- and z-coordinate of PMT.
                
  TTree	      evt                     Event-wise information.
                x0_cm, y0_cm, z0_cm   x-, y- and z-coordinates of event. 
                E0_MeV                Primary kinetic particle energy. 
                px0, py0, pz0         x-, y-, z-components of primary particle momentum
                                        normalized to 1. 
                edepScint_MeV         Energy deposition in scintillator. 
                edepHolder_MeV        Energy deposition in scintillator holder.
                edepWater_MeV         Energy deposition in outer tank filler.
                nPMTs                 Number of fired PMTs.
                nHits[nPMTs]          Vector of hits in fired PMTs.
                fHitTime_ns[nPMTs]    Vector of first hit times in fired PMTs.
                totalPE               Total number of detected photo-electrons.
                hc_x_cm, hc_y_cm, hc_z_cm
                                      x-, y-, and z-coordinates of hit center.
                nPhotons              Number of photons produced by scintillator.
                phEnergy_eV[nPhotons] Vector of photon energies (*)
                phTrackLen_cm[nPhotons] 
                                      Vector of photon track lengths (*)
                phGenChannel[nPhotons]
                                      Vector of photon generation channels (*)
                phAbsChannel[nPhotons]
                                      Vector of photon absorption channels (*)
                phHitLocalTime_ns[nPhotons]
                                      Vector of photon local hit times (*)
                phHitGlobalTime_ns[nPhotons]
                                      Vector of photon global hit times (*)
                absScint              Number of photons absorbed in scintillator.
                absHolder             Number of photons absorbed in scintillator holder.
                absWater              Number of photons absorbed in outer tank filler.
                absPmtGlass           Number of photons absorbed in PMT glass.
                absPmtEnd             Number of photons absorbed in PMT end.
                absPmtCoating         Number of photons absorbed on PMT coating surface.
                absPhotocath          Number of photons absorbed in PMT photo-cathode.
                absTank               Number of photons absorbed on outer tank surface.
                absConcentr           Number of photons absorbed on concentrator surface.
                
  TGraph        scintSpec_eV, scintSpec_nm    Scintillator emission spectrum.  
  TGraph        scintAbs_eV, scintAbs_nm      Scintillator absorption spectrum. 
  TGraph        acrylAbs_eV, acrylAbs_nm      Scintillator holder absorption spectrum. 
  TGraph        waterAbs_eV, waterAbs_nm      Outer tank filler absorption spectrum.
  TGraph        pmtQE_eV, pmtQE_nm            PMT QE curve    
 -----------------------------------------------------------------------------------------            
\end{verbatim}
\normalsize

\end{appendices}

\end{document}